\documentclass[a4paper,11pt]{article}
\usepackage{pos}

\title{The gravitational wave and radio signals of ultralight dark matter
}

\author*[a]{Fa Peng Huang}

\affiliation[a]{MOE Key Laboratory of TianQin Mission, TianQin Research Center for
Gravitational Physics \& School of Physics and Astronomy, Frontiers
Science Center for TianQin, Gravitational Wave Research Center of CNSA, 
Sun Yat-sen University (Zhuhai Campus), Zhuhai 519082, China}

\emailAdd{huangfp8@sysu.edu.cn}

\abstract{We study the new approaches to explore the ultralight (axion) dark matter by gravitational wave experiments and radio telescope based on the superradiance process around Kerr black hole and the resonant conversion process in the magnetosphere of neutron star or the Sun.}

\FullConference{%
  42nd International Conference on High Energy Physics (ICHEP2024)\\
  17-24 July 2024\\
  Prague, Czech Republic
}

\begin{document}
\maketitle

\section{Introduction}
 For decades, physicists have conducted a variety of experiments in an effort to detect the nature of dark matter (DM). However, the anticipated DM signals have yet to be observed. In particular, both direct detection experiments and collider experiments have failed to detect signals of WIMP DM particles. This situation may point us towards ultralight or heavy DM with new approaches, such as radio telescope (SKA/FAST…) and gravtiational wave (GW) detector (LISA, TianQin, Taiji…).
We focus on the discussion of the GW and radio signals of the ultralight DM. 
Ultralight axion is a promising DM candidate.
In particle physics, axion particle can provide a natural solution for the strong CP problem. And axion-like particles are common predictions in string theory.
When ultralight axion meets the strong magnetic field in the magnetosphere of astrophysical objects, resonant conversion might happen due the thermal mass of photon in the plasma.
Then detectable radio signals can be efficiently produced~\cite{Huang:2018lxq,Buckley:2020fmh,An:2020jmf}.
When ultralight axion meets Kerr black hole, axion cloud can be formed around Kerr black hole through the superradiance process.
The axion cloud and the Kerr black hole can form the so-called gravitational atom (GA). Then abundant GW signals can be produced.
Future GW experiments could provide a unique new approach to explore the ultralight axion or axion DM~\cite{Huang:2018lxq,Buckley:2020fmh,An:2020jmf, Yang:2023vwm, Xie:2022uvp,Yang:2023aak,Yang:2024npd}
\begin{figure}
\begin{center}
\includegraphics[scale=0.48]{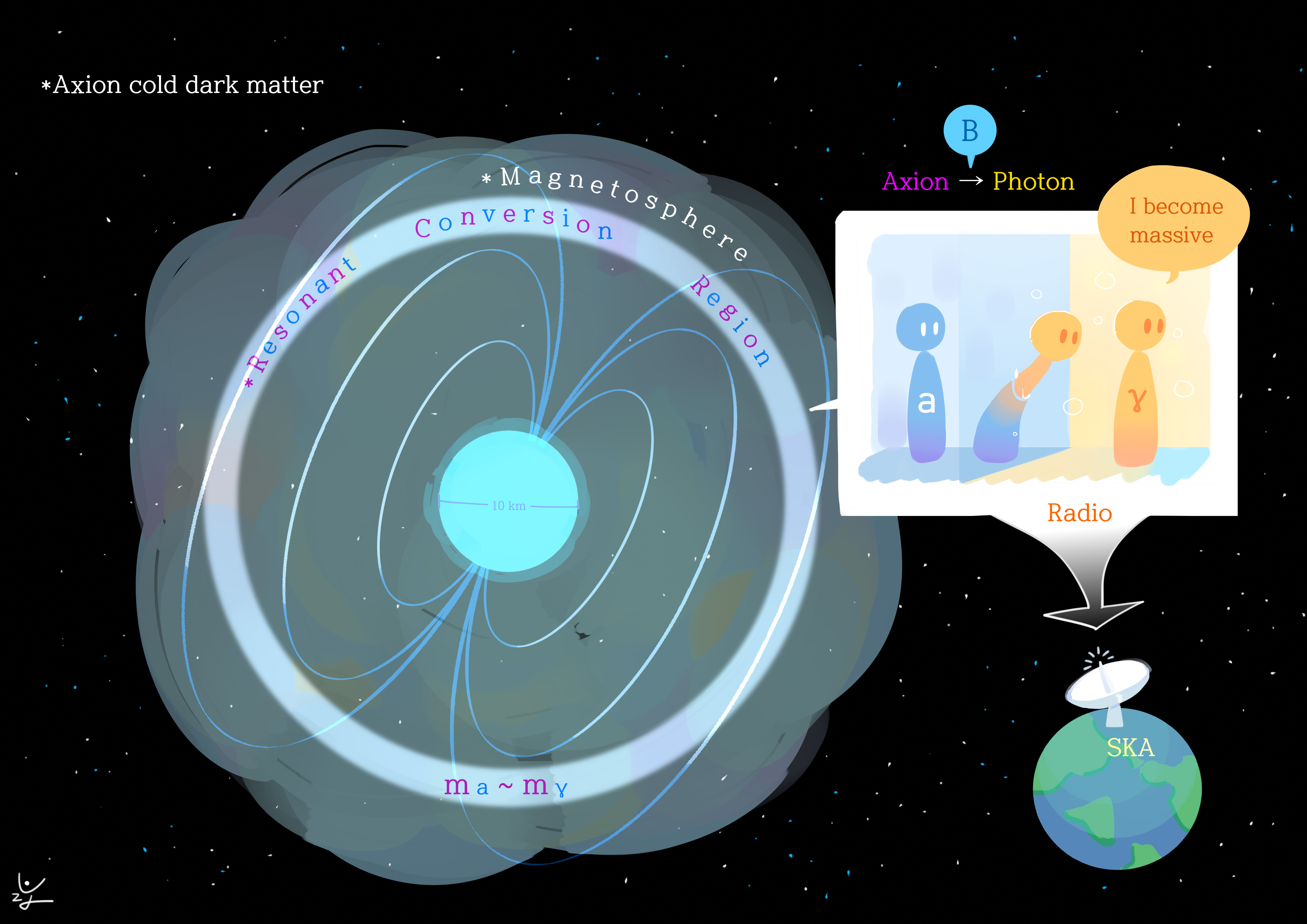}
\caption{The schematic process for the resonant conversion of axion DM to radio signals in the magnetosphere of neutron star.}
\label{fig:axion_neutron}
\end{center}
\end{figure}

\section{ $\mu$eV axion DM and its radio signals}

Recently, in Ref.~\cite{Huang:2018lxq}, our study find the resonant conversion of axion DM to photon in the neutron star magnetosphere can produce strong radio signals. This idea provide new approaches to detect the $\mu$eV axion. 
The basic idea is shown in Fig.~\ref{fig:axion_neutron}.
In the magnetosphere of the neutron 
star, photon obtains effective mass in the plasma. Thus, the photon mass is location 
dependent, and within some region the photon mass approaches the axion DM mass. In the case, resonant conversion happens and the conversion probability is greatly enhanced.
Line-like radio signal for non-relativistic axion might be observed at various radio telescope.
SKA-like experiment can probe the axion DM and the axion mass 
which corresponds to peak frequency.
Working in progress on more delicate study.

Then, we generalize the idea to axion star.
As bosons, axions could condense to axion star and resonantly convert to Fast radio bursts in the magnetosphere of magnetar~\cite{Buckley:2020fmh}.
This scenario provide a natural explaination of the mysterious Fast radio bursts.

\begin{figure}
\begin{center}
\includegraphics[scale=0.27]{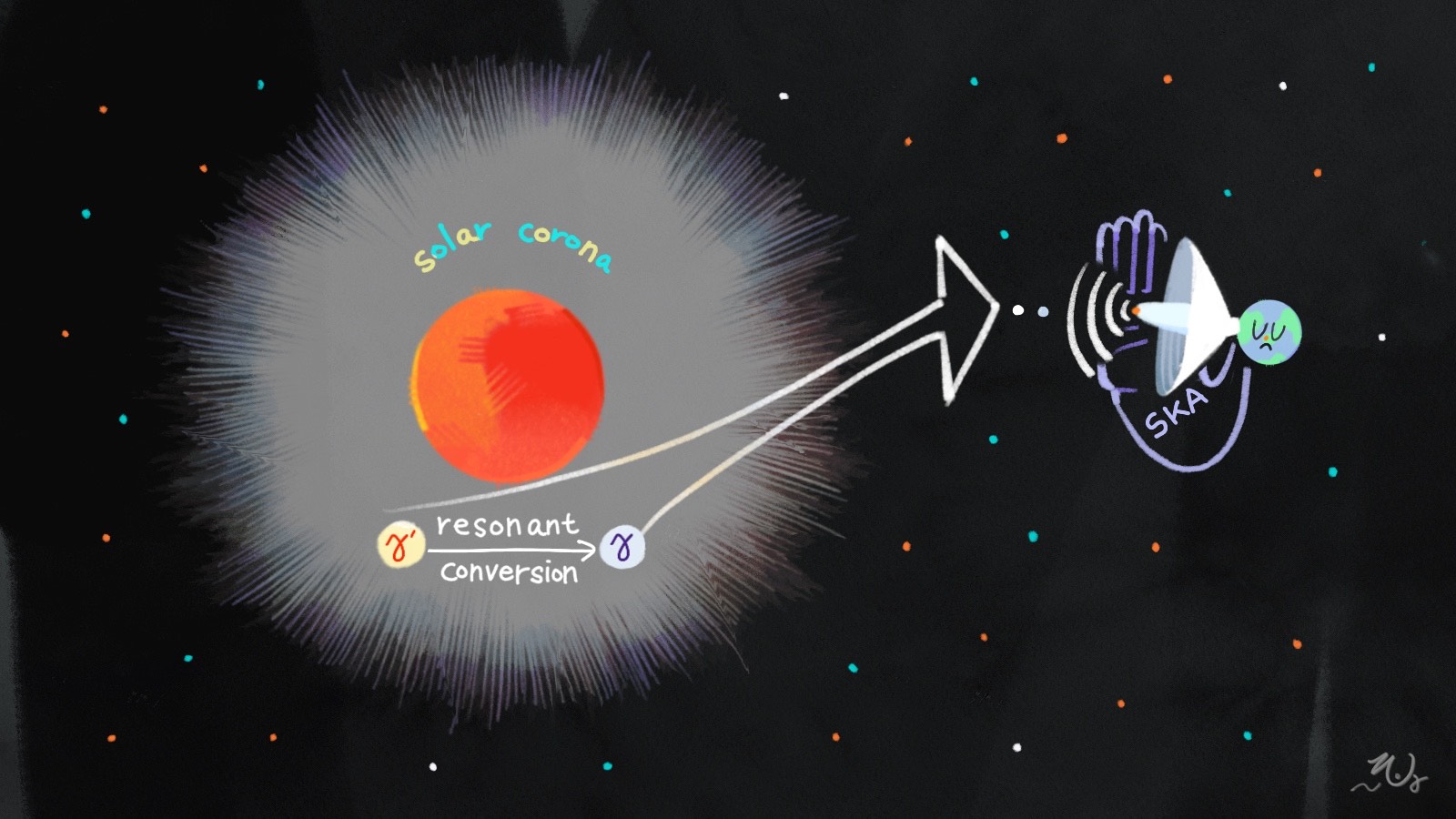}
\caption{The schematic process for the resonant conversion of dark photon DM to radio signals in the solar magnetosphere.}
\label{fig:photon_sun}
\end{center}
\end{figure}

\begin{figure}
\begin{center}
\includegraphics[scale=0.57]{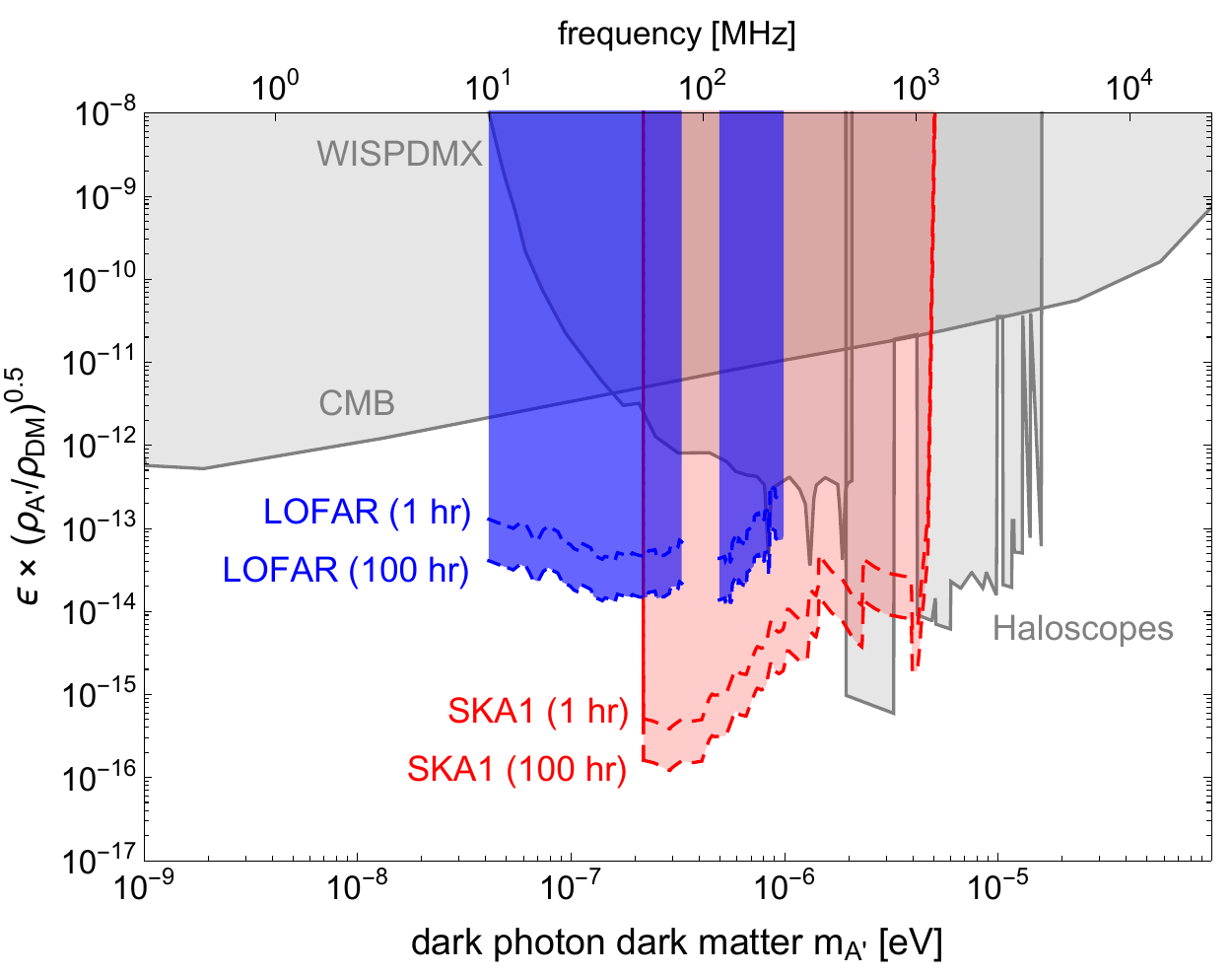}
\caption{The final prediction of the dark photon DM.}
\label{fig:dp_ska}
\end{center}
\end{figure}

We also generalize the axion case to dark photon DM case in the solar magnetosphere~\cite{An:2020jmf}. The resonant conversion process is happened in the solar magnetosphere, as shown in Fig.~\ref{fig:photon_sun}. 
Based on this resonant conversion process, we present the prediction in Fig.~\ref{fig:dp_ska}.

\section{ $10^{-17}-10^{-12}$eV axion DM, GW and pulsar timing measurement}

\begin{figure}[h]
\begin{center}
\includegraphics[scale=0.36]{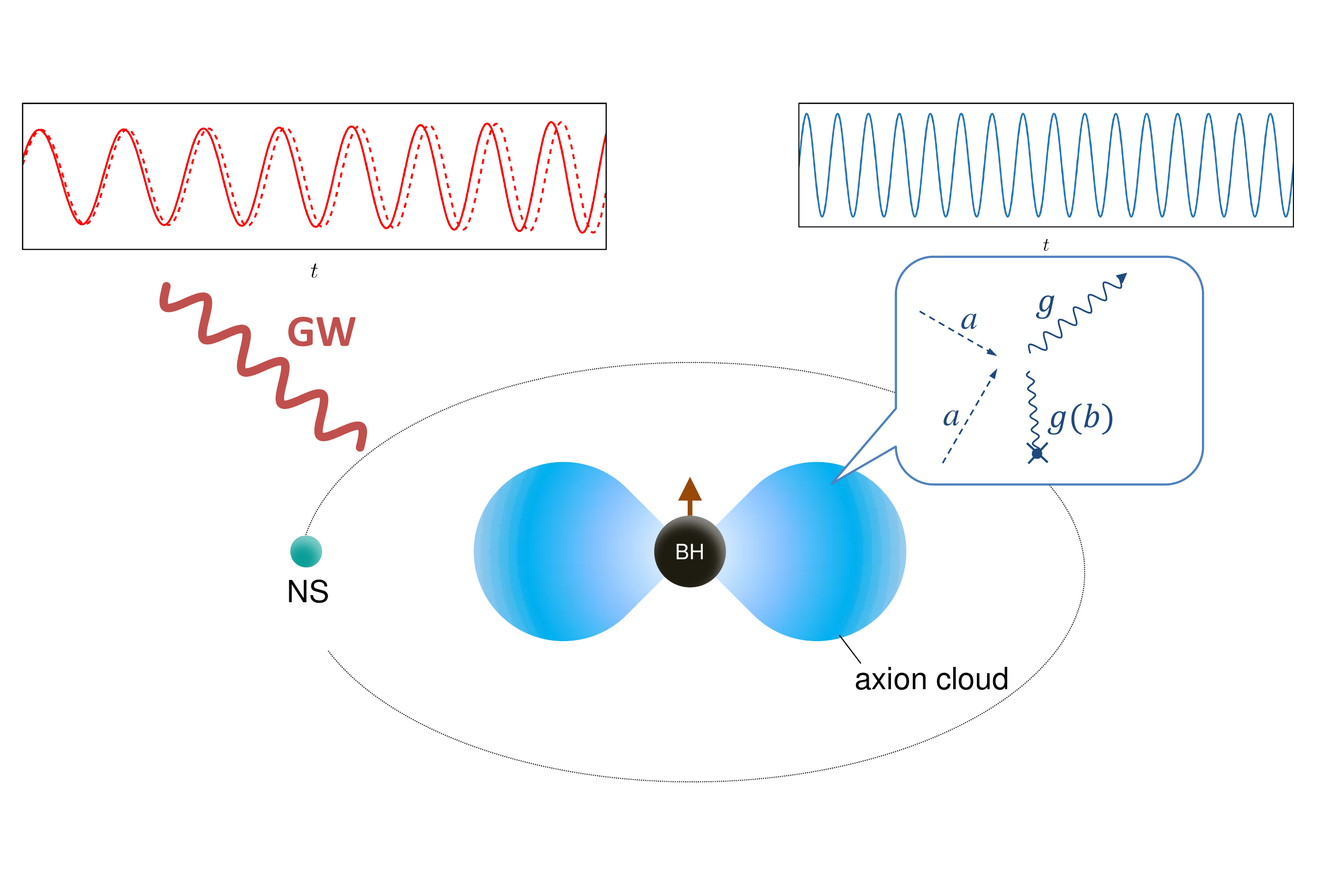}
\caption{The multiband GW radiation for neutron star Kerr blackhole binary with ultralight axion DM.}
\label{fig:sgw}
\end{center}
\end{figure}

When Klein-Gordon meets Kerr, superradiance process may happen with the exponential growth solution of Klein-Gordon equation due to the boundary condition at the horizon of Kerr blackhole. 
Ultralight axion can form axion cloud around rotating BH. GW can be produced from the following two processes: axions can annihilate to GW, energy-level transition of GA. 
Monochromatic GW signal could be produced ~\cite{Yang:2023vwm} and 
gradually depletion of axion cloud could reduce GA mass~\cite{Xie:2022uvp}.
We calculate the GW radiation power through the method of quantum field theory. This method is simple and straightforward compared to previous methods. And it is 
easy to include Kerr metric effects and the 
microscopic physics is intuitive. 
It is clear and simple to demonstrate the analytic approximation formulae~\cite{Yang:2023vwm}. 
The precise prediction is important for the GW and axion search. More precise calculations and more broad applications are working in progress.

\begin{figure}[h]
\begin{center}
\includegraphics[scale=0.98]{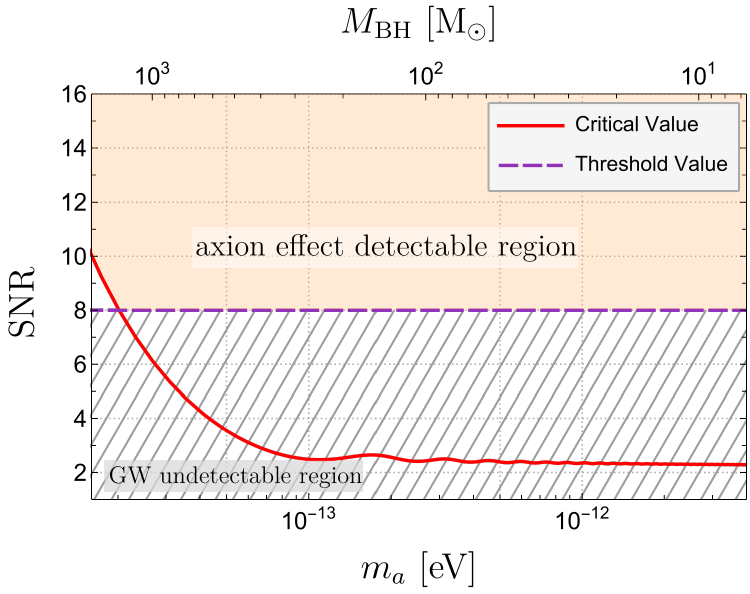}
\caption{The detectability of axion effects at LISA and TianQin.}
\label{fig:snr2}
\end{center}
\end{figure}

We notice the depletion of axion cloud (DC) could reduce GA mass~\cite{Xie:2022uvp} which would lead to obvious phase shift in binary waveform. Taking the neutron star and Kerr black hole binary as an example, we calculate the phase shift and discuss the detectability of ultralight axion effects at LISA and TianQin.
We present the schematic process in Fig.~\ref{fig:sgw}.
The final result is shown in Fig.~\ref{fig:snr2}

\section{$10^{-21}$eV fuzzy axion DM and its GW signals}
Besides the above axion mass regime, the Kerr black hole supperradiance could also provide a new possibility to study the fuzzy axion DM around $10^{-21}$eV.
The cosmic populated supermassive blackhole binaries  dressed with axion cloud as a natural source of nano-Hertz schochastic GW. The energy level transition process can radiate GWs continuously, which naturally fall in nano-Hertz frequency band. 
Consequently, the PTA could detect this new source which provides a new approach to probe ultralight axion DM and isolated blackholes~\cite{Yang:2023aak}.
By Bayesian analysis, we find
fuzzy DM is favored by the data~\cite{Yang:2023aak}.

\section{Conclusion}
Based on the black hole supperradiance process, the resonant conversion process in the magnetosphere of astrophysical objects, and other possible processes, 
GW experiments and radio telescopes might provide new approaches to explore ultralight DM with multi-messenger and multi-band.

\section{Acknowlegments}
This work was supported by the National Natural Science Foundation of China (NNSFC)
under Grant No. 12205387 and No. 12475111.


\begin{thebibliography}{99}

\bibitem{Huang:2018lxq}
F.~P.~Huang, K.~Kadota, T.~Sekiguchi and H.~Tashiro,
Phys. Rev. D \textbf{97} (2018) no.12, 123001
doi:10.1103/PhysRevD.97.123001
[arXiv:1803.08230 [hep-ph]].

\bibitem{Buckley:2020fmh}
J.~H.~Buckley, P.~S.~B.~Dev, F.~Ferrer and F.~P.~Huang,
Phys. Rev. D \textbf{103} (2021) no.4, 043015
doi:10.1103/PhysRevD.103.043015
[arXiv:2004.06486 [astro-ph.HE]].

\bibitem{An:2020jmf}
H.~An, F.~P.~Huang, J.~Liu and W.~Xue,
Phys. Rev. Lett. \textbf{126} (2021) no.18, 181102
doi:10.1103/PhysRevLett.126.181102
[arXiv:2010.15836 [hep-ph]].

\bibitem{Yang:2023vwm}
J.~Yang and F.~P.~Huang,
Phys. Rev. D \textbf{108} (2023) no.10, 103002
doi:10.1103/PhysRevD.108.103002
[arXiv:2306.12375 [hep-ph]].

\bibitem{Xie:2022uvp}
N.~Xie and F.~P.~Huang,
Sci. China Phys. Mech. Astron. \textbf{67} (2024) no.1, 210411
doi:10.1007/s11433-023-2172-7
[arXiv:2207.11145 [hep-ph]].

\bibitem{Yang:2023aak}
J.~Yang, N.~Xie and F.~P.~Huang,
[arXiv:2306.17113 [hep-ph]].


\bibitem{Yang:2024npd}
A.~Yang and F.~P.~Huang,
[arXiv:2404.18703 [hep-ph]].



\end{thebibliography}
\end{document}